# Survey on deep learning in multimodal medical imaging for cancer detection

Yan Tian[1] · Zhaocheng Xu[2] · Yujun Ma[2] · Weiping Ding[3] · Ruili Wang[4] · Zhihong Gao[5,6] · Guohua Cheng[7] · Linyang He[7] · Xuran Zhao[1]

**Abstract**
The task of multimodal cancer detection is to determine the locations and categories of lesions by using different imaging techniques, which is one of the key research methods for cancer diagnosis. Recently, deep learning-based object detection has made significant developments due to its strength in semantic feature extraction and nonlinear function fitting. However, multimodal cancer detection remains challenging due to morphological differences in lesions, interpatient variability, difficulty in annotation, and imaging artifacts. In this survey, we mainly investigate over 150 papers in recent years with respect to multimodal cancer detection using deep learning, with a focus on datasets and solutions to various challenges such as data annotation, variance between classes, small-scale lesions, and occlusion. We also provide an overview of the advantages and drawbacks of each approach. Finally, we discuss the current scope of work and provide directions for the future development of multimodal cancer detection.

**Keywords** Cancer detection · Convolutional neural network · Medical image analysis · Computer vision

## 1 Introduction

Cancer detection [1] in multimodal medical imaging, including X-ray imaging, ultrasonic imaging, whole slide images (WSIs), computed tomography (CT) imaging and magnetic resonance imaging (MRI), has been a promising research area in both academic research and clinical applications. An accurate and efficient cancer detector can be applied to detect various cancers, such as breast, lung,

Yan Tian and Zhaocheng Xu have contributed equally to this work.

This work was supported in part by the National Natural Science Foundation of China under Grant 61972351 and 62111530300; in part by the Special Project for Basic Business Expenses of Zhejiang Provincial Colleges and Universities under Grant JRK22003 and in part by Zhejiang Engineering Research Center of Intelligent Medicine under Grant 2016E10011.

✉ Weiping Ding
  dwp9988@163.com

✉ Ruili Wang
  Prof.ruili.wang@gmail.com

✉ Zhihong Gao
  gzh@wzhospital.cn

[1] School of Computer Science and Technology, Zhejiang Gongshang University, Hangzhou 310018, China

[2] School of Mathematical and Computational Sciences, Massey University, Auckland 0632, New Zealand

[3] School of Information Science and Technology, Nantong University, Nantong 226019, China

[4] School of Data Science and Artificial Intelligence, Wenzhou University of Technology, Wenzhou 325000, China

[5] Department of Big Data in Health Science, The First Affiliated Hospital of Wenzhou Medical University, Wenzhou 325000, China

[6] Zhejiang Engineering Research Center of Intelligent Medicine, Wenzhou 325000, China

[7] Jianpei Technology Co. Ltd, Hangzhou 311200, China

and colon cancer, ultimately saving millions of lives and improving the quality of life for patients [2, 3].

Prior to the era of deep learning, machine learning (ML) techniques were extensively used for cancer diagnosis. For example, support vector machines (SVMs) were used for breast cancer analysis [4, 5], while a decision tree was applied for the analysis and classification of lung cancer morphology [6]. However, with the recent development of cancer detection methods based on convolutional neural networks (CNNs), there has been increasing interest due to their ability to extract local features hierarchically from regularized data [7, 8].

Several surveys have been conducted in this field [9, 10]. Hu et al. [11] provided a comprehensive review of deep learning in cancer detection and diagnosis and proposed future research directions. Rathore et al. [12] discussed various techniques for specific cancer detection tasks and evaluated them using multiple medical datasets. According to Baumgartner et al. [13], model configuration and iterative processes are crucial components in tumor detection, and they proposed a systematic and automated model configuration. Similarly, Saba et al. [14] analyzed and reviewed the current development of tumor detection in various organs using machine learning techniques, including breast, brain, lung, liver, and skin. SkinNet-ENDO [15] employs a deep neural network and an optimization algorithm that uses entropy-normal distribution with ELM (extreme learning machine) to identify skin lesions, with its effectiveness closely tied to the quality of the training data.

However, the aforementioned surveys have primarily focused on categorizing various approaches, while overlooking the practical requirements essential for effectively addressing cancer detection problems in medical imaging. In the context of multimodal medical imaging, there are three primary challenges to cancer detection: (1) The current diagnosis process heavily relies on the expertise of radiologists, leading to a significant workload and reduced efficiency [16–18]; (2) The variance in the appearance of lesions and the high similarity between lesions and the surrounding environment or other organs in medical images make it challenging to accurately identify lesions by deep learning models [19–21]; (3) Due to the difficulty of medical image annotation, only highly professional medical staff can reasonably identify the cancer lesion area, which produces insufficient and incomplete labeled data, resulting in challenges for model training [22, 23].

In this paper, we describe typical challenges that arise in cancer detection, discuss various approaches based on deep learning to handle these challenges, and point out their respective strengths and weaknesses. In addition, we propose a discussion section that aims to analyze the progress made in solving challenges and provide predictions for future studies to continually overcome these challenges.

The remainder of this paper is organized as follows. Section 2 illustrates the background of multimodal medical imaging for cancer detection. In Sect. 3, the survey methodology is described, and the main challenges of deep learning-based cancer detection are listed. Section 4 provides a detailed account of the solutions proposed for the identified problems and challenges. Section 5 presents a comprehensive analysis from an overall perspective and future development direction. Finally, Sect. 6 concludes the paper.

## 2 Background

### 2.1 Medical imaging techniques

There are different imaging techniques for medical applications. X-ray photography is the primary tool and is widely used in tumor diagnosis for chest and abdomen photography due to its high speed and low cost [24–27]. CT scans are often preferred for chest and abdomen examinations because their sensitivity and accuracy are superior to those of X-ray imaging for medical diagnosis in these regions [28–30]. Additionally, chest CT is essential for certain diseases, such as pulmonary fibrosis, during tumor curing. Other imaging techniques, such as MRI [31, 32] and WSIs [33, 34], are used by hospitals for medical diagnosis in specific situations.

### 2.2 Public datasets

Public datasets play a crucial role in measuring and evaluating medical image analysis approaches, promoting the development of the field. In this paper, we focus on object detection tasks for specific tumors such as lung nodules and breast tumors and report several popular datasets. The comparison of multiple datasets is reported in Table 1, in terms of the number of lesion categories, number of medical images, data type, and lesion type. Moreover, there are specific datasets on other cancers or diseases, and more details can be seen in the work [13, 14].

### 2.3 Cancer detection issues

Cancer detection aims to automatically localize and classify potential lesions represented by bounding boxes in medical images. The current research in cancer detection focuses on multiple aspects, as illustrated in Fig. 1. Object detection methods can be classified into two categories based on whether proposals are generated in an early stage. The first category, known as one-stage methods, predicts

Table 1 The statistics of different datasets in cancer detection

| Dataset | Category | Number of images | Data type | Lesion type |
| --- | --- | --- | --- | --- |
| DeepLesion | 8 | 32,120 | CT | Lung/abdomen/kidney |
| Chest X-ray | 14 | 112,120 | X-ray | Chest |
| LUNA16 | 1 | 888 | CT | Lung |
| LIDC-IDRI | 1 | 1018 | CT/X-ray | Lung |
| INbreast | 4 | 410 | X-ray | Breast |
| CAMELYON16 | 3 | 399 | WSI | Breast |
| IDRiD | 4 | 516 | RGB | Fundus |

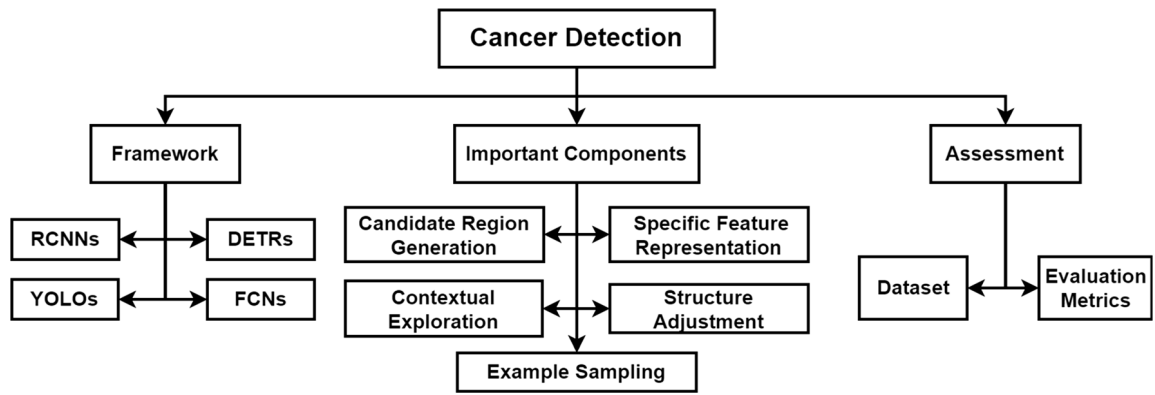

Fig. 1 Current research interests in cancer detection, which are focused on designing detection frameworks that balance effectiveness and efficiency, developing important components to improve performance, and conducting assessments. The term RCNNs refer to region-based convolutional neural networks, while YOLOs are used to describe object detectors from the YOLO family. DETRs, on the other hand, refer to detection transformers, and FCNs stand for fully convolutional networks

bounding boxes in one step with anchors of predefined sizes at predefined locations, such as YOLO [35] and SSD [36]. The second category, known as two-stage methods, is based on region proposals, such as Faster R-CNN [37]. In either case, the quality of the predicted results can be evaluated using the intersection over union (IoU) metric [38]. In addition, fully convolutional networks [39] and transformer-based architectures have also received increasing interest in recent years. Several crucial components of a detector have been extensively studied, such as training example sampling [28, 40], context exploration [41–43], and structural adjustment [44].

## 3 Methodology

We conducted a comprehensive review of more than 150 papers, primarily sourced from leading journals and conferences in the medical field, such as the International Conferences on Medical Image Computing and Computer-assisted Intervention (MICCAI), Medical Image Analysis (MIA), and IEEE Transactions on Medical Imaging (TMI). Other studies published in other journals with high citations were also included. The source distribution of the reviewed papers is illustrated in Fig. 2. The five most highly cited papers in our survey since 2019 are reported in Table 2.

Deep learning and medical image analysis have developed rapidly in recent years. To identify new and innovative approaches for deep learning-based cancer detection, we conducted a thorough review of recent literature. The temporal distribution of the reviewed papers is shown in Fig. 3.

According to these recent works, several challenges have been identified and analyzed in deep learning-based cancer detection:

1. Annotation: How can data annotation be performed on the condition that experienced experts are unavailable or scarce? How can the most valuable samples be selected for annotation when abundant data are collected?
2. Small-scale lesion and occlusion: Crucial information can be lost during encoding owing to lesions in tiny areas or partial occlusion by other organs in the imaging. X-ray images can present unique challenges as images of different organs can interact with each other due to the perspective effect.
3. Small variance of interclass: The issue arises due to the resemblance between lesions in the foreground and organs in the background of the human body.

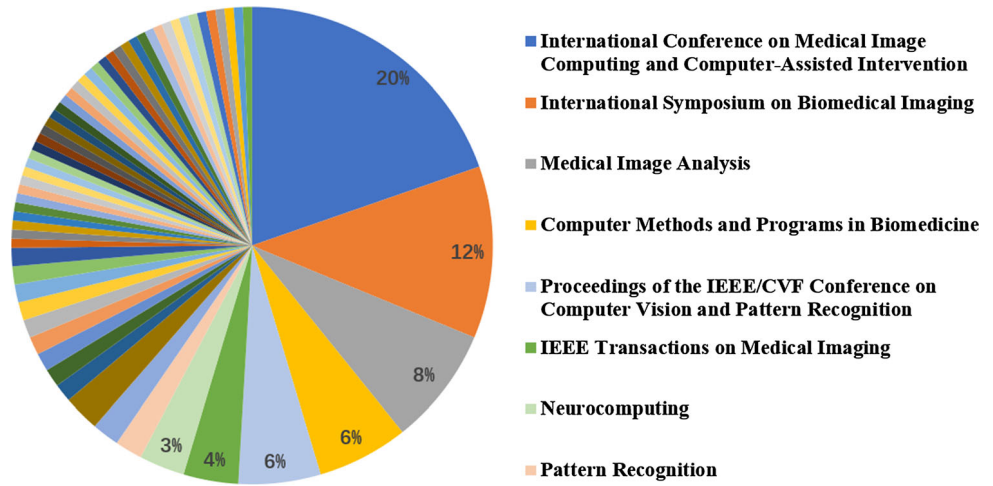

**Fig. 2** Journal/conference distribution of the reviewed contributions

**Table 2** Most highly cited papers in our survey since 2019

| Title | Reference | Journal/conference | Year |
|---|---|---|---|
| Automated pulmonary nodule detection in CT images using deep convolutional neural networks | Xie et al. [45] | Pattern recognition | 2019 |
| Application of convolutional neural network in the diagnosis of the invasion depth of gastric cancer-based on conventional endoscopy | Zhu et al. [46] | Gastrointestinal endoscopy | 2019 |
| Dual-stream multiple instance learning network for whole slide image classification with self-supervised contrastive learning | Li et al. [47] | CVPR | 2021 |
| Recent advancement in cancer detection using machine learning: systematic survey of decades, comparisons and challenges | Saba et al. [14] | Journal of infection and public health | 2020 |
| Robust breast cancer detection in mammography and digital breast tomosynthesis using an annotation-efficient deep learning approach | Lotter et al. [48] | Nature medicine | 2021 |

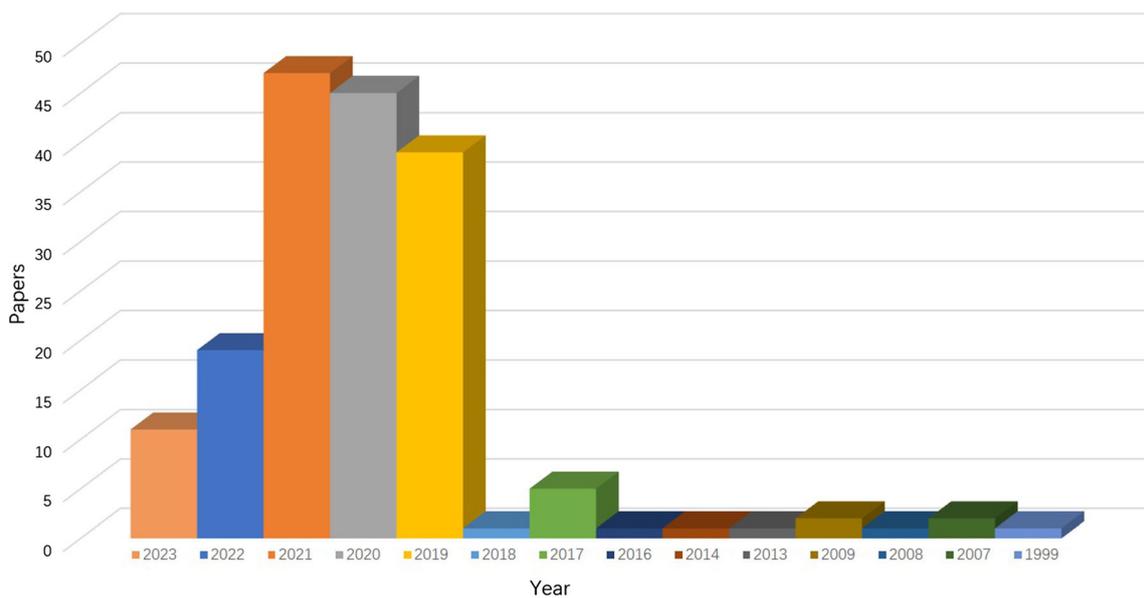

**Fig. 3** Temporal distribution of the reviewed contributions

4. Multitask learning: A unified framework was developed to incorporate other tasks, including segmentation and classification, to aid in detection tasks. However, in some cases, these additional tasks may disrupt rather than facilitate the primary detection task.
5. Generalization: The effectiveness of new data that have diversity is crucial in real applications, which highlights the fact that an overfitted model lacks robustness.

# 4 Main investigation

In contrast to previous reviews, we focus on challenges in deep learning-based cancer detection. We present recent works organized according to the specific problems they aim to address. The challenges and corresponding solutions described in this survey are depicted in Fig. 4.

## 4.1 Framework challenges

The inherent structural differences among detectors present certain challenges in their application for multimodal cancer detection. Aly et al. [49] compared the YOLO architecture with ResNet and Inception and demonstrated that architectural differences have an impact on performance in digital mammogram detection. To fully utilize the feature map information from the SSD feature pyramid and achieve higher accuracy, Zhang et al. [50] reused the discarded information from the max pooling layer as additional feature maps to aid in classification and detection. Addressing the limitations of anchor settings in one-stage detectors, Zlocha et al. [51] employed differential evolution algorithms to optimize anchor configurations and improve the detection capabilities for similar lesions in CT images. To address appearance variations, the spatial region proposal network (SRPN) [52] utilizes a similarity-based mechanism with additional custom convolutional layers to enhance the network's representation learning capabilities. To overcome the limitations imposed by the model structure, multitask learning approaches are employed [31, 53], leveraging joint optimization among different subtask branches to break through performance bottlenecks.

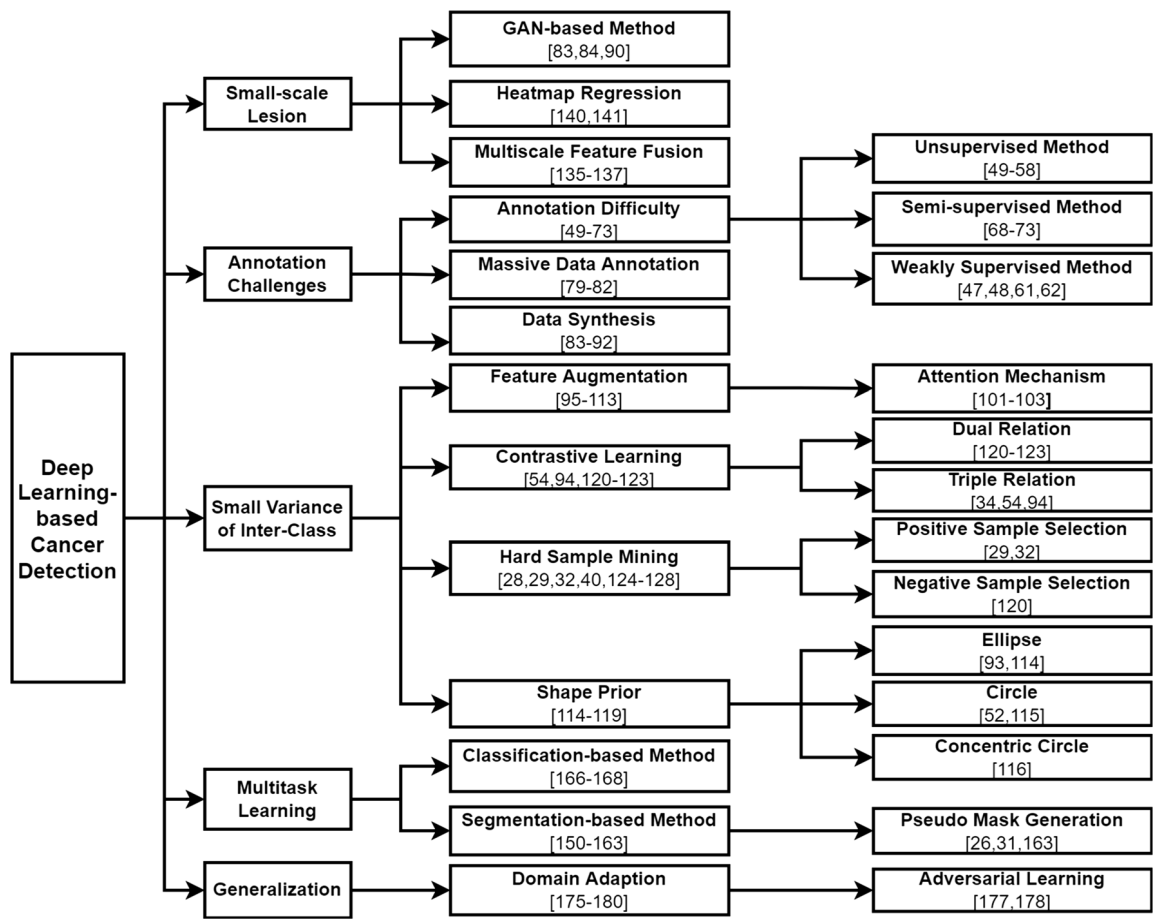

**Fig. 4** An illustration of challenges and current solutions in cancer detection

## 4.2 Annotation challenges

Data-driven methods have been widely explored for the purpose of medical image analysis. However, the progress of research has been impeded by several challenges. One of the major obstacles is the limited availability of experienced experts. In addition, hard sample mining and image synthesis also pose significant challenges that need to be addressed to make further progress.

### 4.2.1 Difficult to annotate

Deep learning models heavily rely on large annotated datasets. However, annotating medical images has always been a challenging task because it requires professional knowledge in the medical field, which makes it difficult to find qualified annotators. To alleviate this annotation difficulty, unsupervised, weakly supervised, and semisupervised methods have been widely used in medical applications. Several noteworthy research studies are summarized in Table 3.

**Unsupervised Methods**. Unsupervised or self-supervised methods are often employed when labeled data are lacking, as medical annotators may not possess sufficient knowledge and the annotation process can be challenging.

One approach to generating dataset-related anchors in unsupervised learning is through clustering methods such as the K-means algorithm [49, 54–57]. To make the algorithm more adaptive to the distribution of target areas, hierarchical clustering [58] can be employed to progressively measure the distance of semantic features extracted through the backbone network for clustering purposes. To explore valuable information hidden within data, self-supervised learning methods [61, 62, 67–70] have been employed for feature learning using given datasets. Specifically, image occlusion and reconstruction are used to augment the robustness of semantic features [71]. Adversarial-based alignment [60] is employed for unsupervised domain adaptation to align the data in source and target domains. Furthermore, global and local adversarial discriminators [59] are utilized to improve the alignment of feature maps across different domains.

**Semisupervised methods**. Semisupervised learning is a technique that involves exploring the features of unlabeled data by learning from the distribution of labeled data without the need for human intervention. Inductive learning [72] is a type of semisupervised learning that focuses on learning the characteristics of both labeled and unlabeled data. On the other hand, transductive learning [73] is a semisupervised learning approach that infers unknown data by minimizing the total loss of unlabeled data.

Semisupervised learning has garnered increasing attention in cancer detection [74, 75], particularly in 3D lung nodule detection. In the context of medical image detection, several techniques have been proposed to address the issue of incomplete or sparse annotations. For instance, the semisupervised medical image detector (SSMD) [63] employs a partly unlabeled dataset and an adaptive consistency loss to train a lesion detector and regularize localization. The two-phase hybrid learning method [76] infers pseudolabels for unlabeled data and incorporates them into model training to address the incomplete annotation issue in signet ring cell detection. Another approach, box density energy (BDE) [64], corrects miscalculations in the loss function resulting from sparse annotation of pathological datasets by conducting loss-calibration based on prediction densities for misprediction penalization. Additionally, positive-unlabeled learning [77] focuses on exploring unlabeled or unknown positive instances while leaving negative samples in the learning stage.

**Weakly Supervised Methods**. Weakly supervised learning is a learning method that falls between supervised and unsupervised learning. Unlike semisupervised methods that blend labeled and unlabeled data, weakly supervised learning methods rely on different supervision signals with less information that is easy to annotate.

One such weakly supervised learning method is multiple instance learning (MIL) [47], which receives bag-level supervision and analyzes the model at the instance level. By using classification signals as guidance, MIL can improve lesion-wise localization ability [48, 65]. In addition to classification signals, electronic medical records (EMRs) are also useful auxiliary information for cancer detection, providing the number and index of nodules [66].

**Table 3** Main contributions to handling the annotation difficulty

| Type | Method | Description |
| --- | --- | --- |
| Unsupervised | K-means clustering [49, 54–57] | Generate anchors by using dataset-specific k-means clustering |
|  | Hierarchical clustering [58] | Agglomerative nesting cluster prediction box based on distance measurement |
|  | Adversarial-based [59, 60] | Align features in source and target domains via adversarial learning |
| Self-supervised | Registering [61, 62] | Maximize image-level similarity between bilateral inputs |
| Semisupervised | Loss function design [63, 64] | Leverage loss function for unlabeled samples |
| Weakly supervised | Multiple instance learning [47, 65, 66] | Aggregate instance-level prediction to obtain bag-level prediction |

Table 4 Main Proposals for small variance of inter-class

| Reference | Method | Description |
| --- | --- | --- |
| Xu et al. [29] | Hard sample mining | Recognize hard positive samples and discard negative samples to improve training |
| Tao et al. [41] | Feature augmentation | Explore cross-slice contextual information and intraslice spatial information via attention mechanism |
| Wang et al. [91] | Shape prior | Utilize anchor shape prior to constrain foreground region information |
| Han et al. [92] | Contrastive learning | Learn distinctive representation by decreasing the distance of embeddings belonging to two classes |

### 4.2.2 Massive data annotation

In the scenarios described in the last subsection, labeled medical images are scarce, and all the data need to provide valuable information. However, in the case of massive data, active learning [78, 79] can reasonably select training samples that are the most conducive to total data for model training without using human resources. Actually, it is already utilized to handle the difficulty brought by massive data in the medical environment [80, 81].

### 4.2.3 Data synthesis

Manual annotation of medical images may not always be feasible or available, resulting in insufficient data for model training [82, 83]. Generative adversarial networks (GANs) [53, 84–87] have been shown to be effective in generating high-quality synthetic data that have the same distribution as the original data, by optimizing the discriminator to distinguish between real and synthetic images during model training.

Several GAN-based approaches have been proposed to generate synthetic medical images. For example, DetectorGAN [88] performs object insertion to generate small-scale synthetic training samples using original images with randomly generated masks. The multiconditional GAN (MCGAN) [89] utilizes dual conditional discriminators related to the lesion context and appearance to generate synthetic images. CycleGAN [90] is a representative cross-modality image synthesis method that can generate target images and source images bidirectionally.

## 4.3 Small interclass variance

As is widely recognized, accurately and efficiently detecting lesions in medical imaging, such as CT and X-ray imaging, is a critical yet challenging task due to the significant intraclass variance and small interclass variance.

### 4.3.1 Feature augmentation

To enhance the discrimination capacity of context features and handle the high similarity between lesion and background in medical images, the attention mechanism [93–95] has been employed in ElixirNet [30] and AttFPN [96]. Later, dual attention [97] and triple attention [98] were used to explore nonlocal dependency in multiple dimensions simultaneously. Recently, a global–local attention mechanism [31, 99–101] has been proposed to simultaneously explore context features at both the global and local levels. Additionally, SNELM [102] combines the squeezenet and extreme learning machine to achieve accurate COVID-19 recognition. However, it may not be suitable for real-time applications due to expensive computation from feature extraction. Other methods for contextual feature exploration, such as feature augmenter [103] and reinforcement learning [43], have also been employed.

To further improve feature learning based on global information, category-specific global prototype alignment [104] is designed to iteratively enhance the compactness of intraclass features based on distance metrics. Additionally, in abdominal organ detection, feature guidance [105] is utilized to increase the interpretability of feature maps. Furthermore, contralateral context information [25] is proposed to explore the structural information of the chest by utilizing patchwise spatial transformers. Multiview image fusion is another approach that builds cross-view latent relations [50, 106–108], typically through image matching techniques such as lesion matching [109, 110] or bipartite graph matching [111], between paired images.

### 4.3.2 Shape prior

Employing the characteristics of biological cells in WSIs, the shape prior of an anchor is a useful strategy for constraining the information coming from the foreground region. The anchor shape prior can take on different shapes, such as an ellipse [91, 112], circle [113], sphere [114, 115], or adaptive shape [116], and it is obtained through data clustering [116, 117] or parameter regression [112]. For instance, circleNet [114] creates a rotation invariant ball-shaped circle representation for anchor-free glomeruli detection. SCPM-Net [115] is designed for 3D center-point representation without pre-defined anchor parameters, using spherical prediction candidates to output the centroid, radius, and offset while employing positive sample selection and matching.

Furthermore, edge cues serve as a valuable source of low-level information for exploring detailed insights. For instance, the edge perceptron [118] extracts and combines edge cues to enhance liver cancer detection in abdominal medical images. Other techniques such as edge dissimilarity measurement [84] and lesion contour searching [119] are also utilized to provide more detailed information.

### 4.3.3 Contrastive learning

To address the challenge of cancer detection, some experts have employed contrastive learning as a means of improving representation learning. They introduce an additional regularization term to the learning process, which enhances the quality of the embeddings. The dual loss [52, 92, 120–123] has been a popular choice for this purpose, which increases the distance between feature maps belonging to two different classes. Alternatively, the triplet loss [34, 52] uses feature maps from three classes. Contrastive-induced gated attention (CIGA) [123] is an example of this approach, which leverages normal and abnormal images to learn robust feature representations.

### 4.3.4 Hard sample mining

Hard samples are located close to the classification boundary with classification difficulty. Inspired by the support vector machine (SVM) [124], hard sample mining [29, 32, 125, 126] has been developed to identify and recognize such hard samples, which in turn improves the training of the detector. In regard to lesion detection, all positive samples must be retained for model training, as they are crucial to the process. Therefore, hard example mining mainly focuses on selecting hard negative samples [120] while discarding a large number of easy negative samples through the use of cascaded classifiers [127], classification scores [128], or sampling strategies [28, 40].

### 4.4 Class imbalance

Class imbalance is a commonly encountered issue in detection tasks, where the detector is more focused on the major class, resulting in poor performance for minority class samples. To address this issue, a sampling strategy is often employed to reduce the number of samples in the majority class. One common method is hard example mining, which explicitly identifies and selects ambiguous samples as hard examples for model learning [29, 126, 127, 129]. Another approach is the deep cascade framework, which uses sequences of decision trees to progressively reduce the number of easy background samples [127]. Additionally, some methods explore the relationships between different samples to identify and exclude those with less informative value.

Some approaches, such as the Bayesian framework with normalization mutual information [130, 131] and importance-aware balanced group softmax (IaBGS) [132], utilize a relation module based on appearance and location weighted features to select valuable examples and reduce the number of training samples belonging to the majority class. However, in contrast, some experts attempt to increase the number of training samples for the minority class. To accomplish this, a box-to-map approach [133] is proposed that utilizes a continuous function to increase samples in three different directions and provide pixel-level supervision, resulting in sufficient positive RoIs. In addition, to generate pseudomasks of new samples in the region proposal network (RPN) stage, a linear interpolation of soft label objectness maps is employed [134].

### 4.5 Small-scale lesion detection

Detecting small-scale lesions has always been a challenge in the medical domain. To achieve high-quality lesion detection for small-scale lesions, it is necessary to exploit additional information. For example, multiscale feature fusion, which combines low-level high-resolution features with high-level semantic features [135–137], or multitask learning, which uses morphology segmentation to assist detection [138, 139], is effective methods for improving small-scale lesion detection. Recently, generative models have also played an important role in addressing small-scale data scarcity. Models such as DetectorGAN [88], FSOD-GAN [82], and SOD-GAN [83] have been proposed to generate synthetic images with small-scale lesions inserted, facilitating the training of deep learning models for lesion detection.

Traditional methods for segmenting lesions in medical images rely on annotating small discrete masks within the images, which can miss valuable information in the encoder and fail to account for unclear boundaries of small-scale lesions. To address this, the heatmap regressor [140, 141] employs lesion distributions modeled by a Gaussian function as segmentation supervisions instead of discrete masks. This approach can better recognize and explicitly represent unclear boundaries of small-scale lesions [142]. However, to further improve the representation of small lesions, size adaptive bounding maps [134] use area-related hyperparameters to adapt to different lesion sizes. Another related method is IoU self-normalization [143], which mitigates the negative effects of loss values influenced by misleading classification results and improves the recall of small nodules.

### 4.6 Occlusion

Occlusion remains a problem in some medical images, such as X-ray imaging. However, the nature of this problem may differ from that in general cases due to the potential interaction between different organs in the imaging process. To address this challenge, traditional methods have utilized data purification [85] or shape priors [144] to detect partially occluded objects. Another approach, CXray-EffDet [145], uses a two-stage pipeline to detect the presence of chest diseases. While effective for detecting known chest diseases, its performance is limited for certain types of unseen chest diseases.

A high-to-low multilevel method [146] has been proposed to address the occlusion problem by considering the continuous distribution of the intensity of mass. However, the context information under severe occlusion can negatively impact the multilevel analysis. To address this issue, CompositionalNets [147] has been proposed to disentangle the representation of the context and foreground object with a part-based voting scheme to match the finer points of the object. Additionally, to alleviate severe occlusions of overlapped chromosomes, template module [148] has been designed to introduce isolated general-template masks with corresponding embeddings representing geometric patterns of chromosomes.

### 4.7 Multitask learning

Multitask learning involves utilizing the interaction among different subtasks to enhance the feature representation within a single model. A popular framework involves separating the shared feature maps into detection and segmentation branches, and simultaneously optimizing the loss of localization and pixel-level classification [42, 53, 149–163]. Another approach involves utilizing an auxiliary task to assist the detection task [84]. A general detection framework for multitask learning is illustrated. The key challenge is constructing pixel-level supervisions for lesion segmentation, where various techniques such as RECIST annotations [26, 31, 51] and Gaussian distribution [140, 141, 164] have been employed to generate masks for lesion segmentation.

To enhance the accuracy of medical image analysis, instance segmentation and semantic segmentation [99] can be used together to incorporate instance-level and semantic-level cues. Additionally, to consider prior knowledge of tumor boundaries and shapes, a label augmentor [103] has been introduced to integrate segmentation and boundary information by expanding scalar labels into vectors. Hybrid lesion detectors with classifiers [104, 165–168] can leverage image-level auxiliary information to provide class-level signals. To avoid suboptimal solutions that result from conflicting tasks and encourage feature diversification, decoupled feature maps from different decoder stages [169] are employed for different tasks. To further reduce subjective judgment from annotators and ensure consistency between local annotations and image-level supervision, extra image-level classification [170] has been designed to use encoder features to indicate the probability of different classes and assist lesion detection.

### 4.8 Postprocessing

Postprocessing is an essential step in cancer detection, as it refines the output predictions and obtains high-quality results according to predefined criteria. One example of such a criterion is the user-defined threshold based on the intersection over union (IoU) [171] or distance-based metric [172], which can be employed to filter out overlapping candidates and incorrect localizations, thus reducing redundancy and matching each prediction with the ground truth.

In the context of cancer detection, the cascaded reduction framework [138] utilizes a patch-level classifier for the lesion area. Spatial transformation [128] can be applied before the cascaded classifier to align the candidate proposal with the appearance-invariant template. To address the suboptimal issue caused by independent prediction in two-step approaches, cascaded detectors [146, 173] integrate an independent classifier with a lesion detector for joint training in an end-to-end manner. Moreover, to further enhance the temporal consistency, series-level postprocessing [174] has been proposed to integrate detection results from every single frame and reduce false alarms in digital subtraction angiography (DSA).

### 4.9 Generalization

Generalization is a crucial aspect of the predictive ability of a trained model on new data. However, if a model suffers from overfitting, it may not be able to make accurate predictions on unknown data, despite performing well on the training data.

### 4.9.1 Domain adaptation

Domain adaptation, a classical approach in transfer learning, has been widely studied in the context of cancer detection [27, 46, 166, 175, 176]. This approach involves training a cancer detector on a dataset from the source domain, and then using the same detector to achieve excellent performance on a dataset from the target domain with different data distributions. Adversarial learning is often employed to ensure the consistency of feature distributions between the source and target domains. For example, the domain-attentive universal detector [177] utilizes an attention mechanism to learn multidomain and domain-invariant knowledge. While BF2SkNet [178] employs a deep learning and fuzzy entropy slime mold algorithm-based architecture for multiclass skin lesion classification.

To further narrow the gap caused by the data distribution of different datasets, Gaussian Fourier domain adaptation (GFDA) and hierarchical attentive adaptation (HAA) were proposed in [179] for interdomain style alignment and scalable domain-invariant feature learning. Moreover, pseudocell-position heatmaps based on Gaussian distribution have been employed to train detectors in the target domain through an iterative process of obtaining pseudomasks and inferring bounding boxes [180].

## 5 Discussion

Data play an important role in learning-based approaches. However, annotation becomes a challenge in medical imaging analysis owing to the scarcity of experienced experts. Unsupervised, semisupervised, and weakly supervised learning are different ways to handle this issue with distinct effects and limits. Unsupervised learning does not explore knowledge from the annotation; therefore, its effectiveness is constrained. Weakly supervised learning seems to be a helpless choice because strong supervision provides more useful information than weak supervision. Semisupervised learning offers excellent scalability, as it allows for the continuous addition of labeled and unlabeled data to the dataset, progressively enhancing the effectiveness of the detector.

The classification task becomes more challenging in the presence of high intraclass variance or small interclass variance. To address this issue, increasing the diversity of samples and employing feature augmentation have emerged as popular methods. It is important to note that conventional data augmentation techniques like image flip and transition alone may not effectively enhance diversity. In addition, contrastive learning can be utilized to investigate the relationship between samples and emphasize the shared knowledge of tumors across different patients. Recent advancements in the field have introduced attention mechanisms and transformers [93], which enhance the discriminative capacity of the model by directing attention to the distinctive regions of the foreground lesion. However, feature augmentation in medical imaging often introduces computational complexity, as there are reasonable discrepancies between tumor regions and other organ tissues. Some experts have attempted to tackle this challenge by disregarding the negative impact of normal organ tissues with distinct shape priors.

In medical image analysis, there is often a class imbalance issue where the number of tumor regions is significantly smaller than that of normal organ regions. To address this issue, the focal loss technique [181] provides a soft method that dynamically adjusts weights for samples from different classes. However, some alternative methods adopt hard example mining strategies to either reduce the number of training samples from the majority class or increase the number of samples from the minority class. Unfortunately, these approaches often sacrifice flexibility. Additionally, the detection of small-scale or occluded lesions poses a challenge as it results in the loss of valuable and discriminative information. To tackle it, several approaches have been commonly employed to enhance the process of feature learning. These include the utilization of hyperfeature (multiscale feature fusion), multitask learning, and GANs. Furthermore, template matching is revisited and combined with data-driven approaches, with a particular emphasis on highly confident regions.

In future, CNN-based approaches will undoubtedly continue to make significant contributions to medical image analysis given their proficiency in handling regularized data. However, there are specific factors, such as X-ray artifacts, that hinder the effective utilization of convolutional neural networks (CNNs) in cancer detection. For instance, the presence of metal or high-density human tissue can result in sound waves or rays being reflected, leading to image distortions or highlighted areas. These artifacts can cause the pattern of the lesion to deviate from the training data, ultimately compromising the performance of the detector. Additionally, when applying learning-based methods to volume-based data, such as CT imaging, it is important to consider efficiency and memory requirements. To tackle this challenge, techniques like data distillation or depthwise separable convolutions can be employed to compress networks and decrease computational complexity.

# 6 Conclusion

In this paper, we focus on the challenge of detecting tumors in multimodal medical imaging, including issues such as dataset construction, annotation, small variance of inter-class, small-scale lesions, and occlusion. We conduct a thorough analysis of various approaches, examining their strengths and drawbacks. Based on our analysis, we predict that learning-based methods will continue to play a prominent role in future of cancer detection.

**Data availability statements** Data derived from public domain resources.

# Declarations

**Conflict of interest** All authors declare that they have no conflicts of interest regarding the publication of this paper.

Neural Computing and Applications